\documentclass[12pt,aps,floats,showpacs,amssymb,tightenlines]{revtex4}

\usepackage{amsmath}
\usepackage{amsfonts}
\usepackage{amscd}
\usepackage{epsfig}
\usepackage{amssymb}
\usepackage{tabularx}
\def\a{\alpha}

\def\m{\mu}
\def\n{\nu}
\def\d{\delta}

\def\L{\Lambda}

\def\w{\omega}
\def\c{\nabla}
\def\p{\partial}
\def\k{\kappa}
\def\mm{\tilde{\mu}}
\def\ll{\tilde{\Lambda}}

\begin{document}

\title{Linear stability of Einstein-Gauss-Bonnet static spacetimes- Part II: Vector and scalar perturbations}
\author{Reinaldo J. Gleiser and Gustavo Dotti}
\affiliation{Facultad de Matem\'atica, Astronom\'{\i}a y
F\'{\i}sica, Universidad Nacional de C\'ordoba, Ciudad
Universitaria, (5000) C\'ordoba, Argentina}
\email{gdotti@famaf.unc.edu.ar}

\begin{abstract}
 We study the stability under linear perturbations of a class of static
solutions of  Einstein-Gauss-Bonnet gravity in $D=n+2$ dimensions
with spatial slices of the form $\Sigma_{\k}^n \, \times \, {\mathbb
R}^+$, $\Sigma_{\k}^n$ an $n-$manifold of constant curvature $\k$.
Linear perturbations for this class of space-times can be generally
classified into tensor, vector and scalar types. In a previous
paper, tensor perturbations were analyzed. In this paper we study
vector and scalar perturbations. We show that vector perturbations can be analyzed in
 general using an S-deformation approach and do not introduce instabilities. On the other hand,
 we show by analyzing an explicit example that, contrary to what happens in
Einstein gravity, scalar perturbations may lead to instabilities in
black holes with spherical horizons when the Gauss-Bonnet string
corrections are taken into account.
\end{abstract}

\maketitle

\section{Introduction} \label{intro}
The analysis of the properties and behavior of gravity in higher
dimensions has become a major area of research in recent years,
motivated in particular by developments in string theory. Among
others, the Einstein-Gauss-Bonnet (EGB) gravity theory has been
singled out as relevant to the low energy string limit \cite{str1}.
The EGB theory is a special case of Lovelock's theory of gravitation
\cite{Lovelock}, whose lagrangian is a linear combination of Euler
densities continued from lower dimensions. Lovelock's theory gives
equations involving up to second order derivatives of the metric,
and has the same degrees of freedom as ordinary Einstein theory
\cite{Lovelock}. A number of  solutions to the EGB equations, many
of them relevant to the development of the $AdS-CFT$ correspondence
\cite{jm}, are known, among them a variety of black holes in
asymptotically Euclidean or $(A)dS$ spacetimes
\cite{w1,w2,w3,atz,chm,bd}. These were found mostly because they are
highly  symmetric. Analyzing their linear stability, however,
confronts us with the  complexity of the EGB  equations, since the
perturbative terms break the simplifying symmetries of the
background metric. To be more specific, we consider spacetimes that
admit locally a metric of the form
\begin{equation} \label{b}
ds^2 = -f(r) dt^2 + g(r) dr^2 + r^2 \bar g _{ij} dx^i dx^j,
\end{equation}
where $\bar g _{ij} dx^i dx^j$  is the line element of an
$n-$dimensional
 manifold $\Sigma_{\k}^n$ of constant curvature $\k=1,0$ or $-1$.
Linear perturbations around (\ref{b}) can be conveniently
classified, following the scheme proposed in \cite{koda}, into
tensor, vector, and scalar perturbations. The $\k=1$ case
$\Sigma_{1}^n = S^n$ gives, for appropriate $f$ and $g$,
cosmological solutions, as well as higher dimensional Schwarzchild
black holes. The stability of these solutions under tensor
perturbations was studied in \cite{dgl}. A complete classification
of solutions to the EGB equation with line element (\ref{b}) and
$\k=1,0$ or $-1$, together with a case by case stability study under
tensor perturbations appears in \cite{dge}. In this paper we extend
the analysis in \cite{dge} to vector and scalar perturbations. The
methods we employ here are rather different from the analytic
computations carried out in \cite{dgl} and \cite{dge} for arbitrary
dimension. The complexity of the computations involved in the study
of vector and scalar perturbations forced us to develop alternative
approaches. Explicit expressions were worked out for space-time
dimensions $D=n+2 \leq 11$, and their $n-$dependence was then
interpolated. As a result, we obtained expressions which are low
degree ($\leq 4$) polynomials in $n$, correct at least up to the
highest physically interesting space-time dimension ($D=11, n=9$).
Nevertheless, we conjecture that they are correct for arbitrary $n$.
This is because only the lower values of $n$ (up to $n=6$) are
required to obtain the $n-$dependence, and higher values appear as
``predictions". Moreover, it is clear that a purely analytic
derivation of the equations for general $n$ should lead to
expressions of the form presented here, and this gives further
support to our conjecture. The paper is organized as follows. In
section \ref{vsp} we give the general solution of the EGB equations
with metric (\ref{b}), then we introduce  vector and scalar
perturbations following \cite{koda}. In section \ref{vegb} we study
the stability of (\ref{b}) under vector perturbations in EGB
gravity. As was done for tensor perturbations in \cite{dge}, we show
that the problem reduces to obtaining a lower bound for the spectrum
of a Schr\"odinger operator which, in spite of its complexity, turns
out to be an $S-$deformation \cite{koda} of a much simpler operator.
Section \ref{segb} is devoted to scalar perturbations. A step by
step guide for constructing the potential of the associated
Schr\"odinger operator is given, since its general expression for
arbitrary dimensions is extremely long and has a complicated
dependence on the parameters. This makes a
general analysis of the stability problem under scalar type perturbations
practically impossible.
On the other hand, since the procedure given in this paper allows for the explicit
construction of the potential for any particular choice of dimension $n$,
and other parameters of the theory, our results can easily be used in the analysis
of particular classes of situations. As an
example and application, we analyze Schwarzschild-like five dimensional black holes and find a scalar instability of low mass, which
is absent in Einstein gravity. Section \ref{conc} contains the conclusions.

\section{Vector and scalar
perturbations of a class of static spacetimes} \label{vsp}
\noindent
The Einstein-Gauss-Bonnet (EGB) equations in the vacuum case are
given by
\begin{equation} \label{lovelock}
0 = {\cal{G}}_b{}^a \equiv \Lambda {G_{(0)}}_b{}^a + {G_{(1)}}_b{}^a
+ \a {G_{(2)}}_b{}^a,
\end{equation}
where $\Lambda$ is the cosmological constant, ${G_{(0)}}_{ab} =
g_{ab}$ the spacetime metric, $ {G_{(1)}}_{ab} = R_{ab} -\frac{1}{2}
R g_{ab}$ the Einstein tensor and
\begin{multline} \label{g2}
{G_{(2)}}_b{}^a = R_{cb}{}^{de}  R_{de}{}^{ca}  -2 R_d{}^c
R_{cb}{}^{da} -2 R_b{}^c R_c{}^a + R  R_b{}^a
 -\frac{1}{4} \d^a_b \left(
R_{cd}{}^{ef}R_{ef}{}^{cd} - 4 R_c{}^d R_d{}^c + R^2 \right)
\end{multline}
the quadratic  Gauss-Bonnet tensor. These  are
 the first  three in a tower
${G_{(s)}}_b{}^a, s=0,1,2,3,...$ of tensors  of order $s$ in
$R_{ab}{}^{cd}$  given  by Lovelock in \cite{Lovelock}.

Here we  consider static spacetimes satisfying (\ref{lovelock}) with
a metric of the form (\ref{b}). These are foliated by spacelike
hypersurfaces, orthogonal to the time-like Killing vector $\p/\p t$,
 that contain  a submanifold of dimension
$n=D-2$ ($D$ the spacetime dimension) of constant curvature $\k=1,0$
or $-1$, and line element $\bar g_{ij} dx^i dx^j$. The non-zero
components of the Riemann and Ricci tensors, and the Ricci scalar
for
 a metric of the form
(\ref{b}) are given in \cite{dge}.
 Inserting these in (\ref{lovelock}) we find that
(\ref{b}) solves the EGB equation (\ref{lovelock}) if
\cite{w1,w2,w3,atz,dgl,dge}
\begin{equation}\label{f}
\frac{1}{g(r)} = f(r) = \k - r^2 \psi(r)
\end{equation}
and $\psi(r)$ is a solution of
\begin{equation}\label{p}
  P(\psi) \equiv \frac{n(n-1)(n-2)}{4} \psi^2 + \frac{n}{2 \a}
\psi - \frac{\Lambda}{(n+1) \a} = \frac{\mu}{\a r^{n+1}}.
\end{equation}
This implies that,
\begin{equation} \label{f1}
 f(r) = \k +\frac{{r}^{2}}{\a(n-1)(n-2)} \left( 1+\epsilon\,\sqrt {1+
 \frac { 4 \a \left( n-1 \right)  \left( n-2 \right)}{n(n+1)}  \left[
\frac{\m (n+1)}{r^{n+1}} + \Lambda \right] } \; \right),
\end{equation}
where $\epsilon=\pm 1$ picks a root of the quadratic polynomial
$P(\psi)$. The Ricci scalar for this solution is \cite{dge}
\begin{equation} \label{ricci}
R =  (n+2)(n+1)\psi(r)+2r(n+2) \frac{d \psi(r)}{dr}+r^2\frac{d^2
\psi(r)}{dr^2},
\end{equation}
so that there is a singularity whenever $\psi \to \infty$ or $\psi
\to \psi_o$, the stationary point of $P$, since $d \psi/d r \to
\infty$ at this point. We also note \cite{dge} that
 if $\m/\a > 0$ the condition  $f=\k-r^2 \psi > 0$ reduces to
\begin{eqnarray} \label{c1}
\psi &\leq & 0 \;\; \text{ or } \;\; 0< \psi, \; \frac{\m}{\a}
|\psi|^{\frac{n+1}{2}} \leq P(\psi) \;\; \text{ ( if }
 \;\; \k=1 ) \\ \label{c0}
\psi &\leq&  0  \;\; \text{ ( if } \;\; \k=0 )\\
\psi &\leq& 0 \;\; \text{ and  } \;\;\frac{\m}{\a}
|\psi|^{\frac{n+1}{2}}
 \geq P(\psi) \;\; \text{ ( if }
 \;\; \k=-1 ) \label{c-1}
\end{eqnarray}
whereas for   $\m/\a < 0$, $f=\k-r^2 \psi > 0$ is equivalent to
\begin{eqnarray} \label{cc1}
\psi &\leq & 0 \;\; \text{ or } \;\; 0< \psi, \; P(\psi) \leq
\frac{\m}{\a} |\psi|^{\frac{n+1}{2}} \;\; \text{ ( if }
 \;\; \k=1 ) \\ \label{cc0}
\psi &\leq&  0  \;\; \text{ ( if } \;\; \k=0 )\\
\psi &\leq& 0 \;\; \text{ and  } \;\;P(\psi) \geq \frac{\m}{\a}
|\psi|^{\frac{n+1}{2}} \;\; \text{ ( if }
 \;\; \k=-1 ) \label{cc-1}
\end{eqnarray}
In \cite{dge},  the space-times
(\ref{b})-(\ref{f})-(\ref{p})-(\ref{f1}) were classified by studying
the intersections of the curves $P(\psi)$ and $\frac{\m}{\a}
|\psi|^{\frac{n+1}{2}}$.

To study linear perturbations we
 follow the treatment given in \cite{koda}, where it is shown that
an arbitrary perturbation of the metric is a linear combination of
perturbations of the tensor, vector and scalar types. Tensor
perturbations around the EGB vacuum solution
(\ref{b})-(\ref{f})-(\ref{p})-(\ref{f1})
 were studied in \cite{dgl,dge}. We consider now
vector and scalar perturbations. We use $a,b,c,d,...$ as generic
indices, greek indices $\mu,\nu$ refer to $r,t$, whereas
$i,j,k,l,m,...$ are  assumed to take values on $\Sigma_{\k}^n$. A
bar denotes tensors and operators on $\Sigma_{\k}^n$. The
perturbations are of the form
\begin{equation} \label{p1}
g_{ab} \to g_{ab} + h_{ab},
\end{equation}
and indices of $h_{ab}$  are raised using the background metric,
therefore
$\d g^{ab} = - h^{ab}$.

\subsection{Vector perturbations}

A general {\em vector} perturbation is given by,
\begin{equation} \label{pert1}
 h_{\mu\nu}=0\;\,\;\; h_{\mu i}=r f_{\mu} V_i\;\;,\;\; h_{ij} = 2 r^2 H_T V_{ij}
\end{equation}
where $f_{\mu}$, and $H_T$ are functions of $(r,t)$, and $V_i$,
$V_{ij}$ are defined on $\Sigma_{\k}^n$. $V_i$ is a divergence-free
vector harmonic field, satisfying,
\begin{eqnarray} \label{vect1}
(\bar{\triangle} +k_V^2)V_i & = & 0 \nonumber \\
\bar{\c}_i V^i & = & 0
\end{eqnarray}
where $\bar{\triangle} := \bar{\c}_j \bar{\c}^j$ and $\bar{\c}_i$
are,
 respectively, the
Laplacian and covariant  derivative for the metric $\bar g _{ij}$ of
(\ref{b}). The symmetric tensor
\begin{equation} \label{tens1}
V_{ij}= - \frac{1}{2k_V}\left( \bar{\c}_i V_j +  \bar{\c}_j V_i
\right)
\end{equation}
is a harmonic tensor on $\Sigma_{\k}^n$, with the properties,
\begin{equation} \label{tens2}
\left[\bar{\triangle} +k_V^2-(n+1)\k \right]V_{ij} = 0
\end{equation}
\begin{equation} \label{tens3}
V^i{}_i=0\;\;,\;\; \bar{\c}_j V^j{}_i = \frac{k_V^2-(n-1)\k}{2k_V}
V_{i}
\end{equation}
As shown in \cite{koda}, for $k_V^2 \neq (n-1)\k$, the combinations,
\begin{equation} \label{guage1}
F_{\mu} = f_{\mu} + \frac{r}{k_V} \p_{\mu}{H_T}
\end{equation}
are a basis for gauge invariant variables.

\subsection{Scalar perturbations}

{\em Scalar} perturbations are of the form
\begin{equation} \label{sca1}
h_{\m \n} = {\cal{F}}_{\m \n} S, \hspace{1cm} h_{\m i} = r
{\cal{F}}_{\m} S_i, \hspace{1cm} h_{ij} = 2r^2( H_L \, \bar g _{ij}
S + H_T S_{ij} ).
\end{equation}
In (\ref{sca1}), $S$ is a scalar harmonic
\begin{equation} \label{sca2}
(\bar \triangle + k^2_S ) S = 0,
\end{equation}
and one constructs  scalar-type harmonic vectors and tensors
\begin{equation}
S_i = -\frac{1}{k_S}  \bar \c_i S, \hspace{1cm} S_{ij} =
\frac{1}{k_S^2} \bar \c_i \bar \c_j S + \frac{1}{n}  \bar g _{ij} S,
\end{equation}
which satisfy
\begin{equation}
\left[ \bar \triangle + k_S^2-(n-1)\k \right] S_i = 0, \hspace{1cm}
\bar \c_i S^i = k_S S, \hspace{1cm} S_i{}^i = 0,
\end{equation}
and
\begin{equation}
\c_j S_i{} ^j = \frac{(n-1)(k_S^2-n \k)}{n k_S} S_i, \hspace{1cm}
\left[ \bar \triangle + k_S^2 -2n \k \right] S_{ij} = 0.
\end{equation}

Harmonic symmetric tensors of arbitrary rank on $n-$spheres are
constructed in \cite{higu}. In general, we lack  explicit
expressions for harmonic tensors on  arbitrary constant curvature
manifolds, although bounds on the Laplacian spectrum can be derived.

 It can be shown \cite{koda} that, in analogy, and, extending to
higher dimensions the well known Regge-Wheeler gauge for ``even''
perturbations in four space-time dimensions, one can, by an
appropriate coordinate
 transformation, choose a gauge where  ${\cal{F}}_a=0$, and $H_T=0$. This makes
 ${\cal{F}}_{ab}$ and $H_L$ a basis for gauge invariant quantities. We shall make
 this choice of gauge in what follows, but, for convenience, we rewrite ${\cal
 F}_{ab}$ in the form,
\begin{equation}\label{sgauge2}
{\cal F}_{rr}= \dfrac{1}{f}F_{rr} \;, \;\;\;{\cal
F}_{rt}=\frac{\partial F_{rt} }{\partial t} \;, \;\;\;{\cal
F}_{tt}={f}F_{tt}
\end{equation}

\section{Vector perturbations in EGB gravity}
\label{vegb}

In this section we study vector perturbations of (\ref{b}) as
defined in equations (\ref{pert1})-(\ref{guage1}).
 As is clear from the
derivations in \cite{koda}, one can always choose a gauge where
$H_T=0$. This generalizes the  Regge-Wheeler gauge for odd
perturbations to higher dimensions, but more important, it
simplifies considerably the analysis. With this
 simplifying gauge choice, after a long computation we find
that the non-trivial linearized Einstein-Gauss-Bonnet equations
$\delta G_{ij}=0$
 are equivalent
to,
\begin{eqnarray}\label{eqij}
0 & = & {\frac {\partial }{\partial r}}\left[F_r\,f\left(r^{(n - 2)}
+ \a\,(n - 2)\,{\frac {d }{d
r}}\left(r^{(n - 3)}\,(\k-f)\right)\right)\right] \nonumber \\
& & -   {\frac {\partial }{\partial t}}\, \left[
   \frac {F_t}{f} \left(r^{(n - 2)}
 + \a\,(n - 2)\,{\frac {d }{d r}}\,(r^{(n - 3)}
\,(\k-f)\right)
 \right]
\end{eqnarray}
while $\delta G_{ir}=0$ can be written as

\begin{eqnarray}
\label{eqirold} 0 & = & r\,\left({\frac {\partial ^{2}F_t}{\partial
t\,
\partial r}}\,r - {\frac {\partial ^{2} F_r
}{\partial t^{2}}}\,r - 2\,{\frac {
\partial F_t}{\partial t}}\right)
 (r^{2} + \a (n - 1)\,(n - 2)(\k - f))\nonumber \\
\mbox{} & & + F_r\,f\left\{
  \left[(n - 1)\,(n
 - 2)\,r^{2}\,(\k - f) - 2\,(n - 1)\,r^{3}\,
{\frac {df}{dr}}-r^{4}\,{\frac {d^{2}f}{dr^{2}}} \right] \right. \nonumber \\
& & + (n - 1)\,(n - 2)\left[ r^{2}\,{\frac {d}{dr}}\,\left({\frac
{df}{dr}}\, (f-\k)\right) + 2\,(n - 3)\,r\,{\frac {df}{dr}}\,
\,(f - \k ) \right. \nonumber \\
& &
\left. + \frac{1}{2}(n - 3)\,(n - 4)\,(f - \k )^{2}\right] \a \nonumber \\
& &
 +\left. \left[\a(n - 2)\,\left((n - 3)\,(f - \k ) + r\,
\frac {df}{dr}\right) - \,r^{2}\right]\,\left( k_V^2-(n-1)\k \right)
- 2 \,r^{4}\Lambda \right\}
\end{eqnarray}
and $\delta G_{it}=0$ leads to,

\begin{eqnarray}
 \label{eqit}
 0 & = & F_t \left\{ - r^{2}\,\left[r^{2}\,
{\frac {d^{2}f}{dr^{2}}} + (n - 1)\,\left(2\,r\, {\frac {df}{dr}} +
n\,(f - \k) + 2\right)\right] \right.
  \nonumber \\
& & +(n-1)(n - 2)\left[r^{2}\,{\frac {d}{dr}}\,\left({\frac
{df}{dr}}\,(f -
\k)\right) + 2\,(n - 2)\,r\,{\frac {df}{dr}}\,f   \right. \nonumber \\
& & + \left. (n - 3)\,\left(\frac{n}{2}\,(\k - f)^{2} - 2 - 2\,r\,
{\frac {df}{dr}} + 2\,f\right)\right]\a - 2
\,r^{4}\Lambda  \nonumber \\
& & -\left. \left[r^{2} - \a \,(n - 2)\,\left(r\,{\frac {df}{dr}} -
(n - 3)\,(\k - f)\right)\right]\,\left( k_V^2-(n-1)\k \right)
\right\}
 \nonumber \\
& & + f\,r^{(4 - n)}\,\left\{ r^2{\frac {\partial }{
\partial r}}\,\left({\frac {\partial F_t}{\partial r}}\,r^{(n - 4)}\,(r^{2} +
\a\,(n - 1)\,(n - 2)\,(\k-f))\right) \right. \nonumber \\
& & -\left. {\frac {\partial }{
\partial r}}\,\left({\frac {\partial F_r }{\partial t}}\,r^{(n - 2)}\,(r^{2}
+ \a\,(n - 1)\,(n - 2)\,(\k-f))\right) \right\}
\end{eqnarray}
As explained in the Introduction, the $n$ dependence in these
formulas was obtained by interpolating the results for different $n$
values, and checked in the physically relevant range $n \leq 9$ and
isolated higher $n$ values. No more than four points were required
for the interpolation in every case. We conjecture that these
formulas are valid for every $n$.  As required for consistency,
(\ref{eqij}), (\ref{eqirold}), and (\ref{eqit}) are not independent.
Setting
\begin{equation} \label{guage2}
F_{t} = p(r) f(r) \frac{\partial \chi(t,r)}{\partial r} \;\;,\;\;
F_{r}=\frac{p(r)}{f(r)}\frac{\partial \chi(t,r)}{\partial t},
\end{equation}
with
\begin{equation} \label{guage3}
 p(r)= \left[ r^{(n-2)} - \a (n-2) \frac{\p}{\p r} \left[
r^{(n-3)}(f(r)-\k) \right] \right] ^{-1},
\end{equation}
solves trivially (\ref{eqij}) and makes both (\ref{eqirold}) and
(\ref{eqit}) equivalent to
\begin{equation} \label{eqRW1}
 \frac{\partial^2 \Phi(t,r)}{\partial t^2}-f^2\frac{\partial^2
\Phi(t,r)}{\partial r^2} +Q_V(r)\frac{\partial \Phi(t,r)}{\partial
r} +q_V(r) \Phi(t,r)=0.
\end{equation}
Here,
\begin{equation} \label{eqRW2-1}
 \Phi(t,r) = \frac{\partial \chi(t,r)}{\partial t},
\end{equation}
and
\begin{equation}
q_V(r) = \frac{f \left( k_V^2 - (n-1)\k \right) H}{r^2}, \label{q1}
\end{equation}
where
\begin{equation} \label{q2}
H = \frac {{\alpha}^{2}n \left( n-2 \right) ^{2} \left( n-3 \right)
\left( n-1 \right) {\psi}^{2}+2\,\alpha\,n \left( n-2 \right)
 \left( n-3 \right) \psi+4\, \left( n-2 \right) \Lambda\,\alpha+2\,n}{
2 n \left(\left( n-1 \right)  \left( n-2 \right) \psi\,\alpha+1
 \right) ^{2}}.
\end{equation}
Also
\begin{multline}\label{Qden}
 Q_V(r) =  - {\frac {  f }{ \left[ \a \, ( n-2
 )  r\left(  ( n-3 ) ( \k-f) -r{\displaystyle{{\frac {df}{dr}}}}  \right) +{r}^{3}
 \right] }}  \\
 \times \left\{ \a ( n-2 )  \left[  r^2 f {\frac {d^{2}f}{d{r}^{2}}} - r^2\left( {\frac {
df}{dr}}  \right) ^{2} -f ( n-2)( n-3 ) (\k-f)
 \right.   \right.
\\
\left. \left.+ \left( { \frac{df}{dr}}  \right)  \left( ( n-1) f
+n-3 \right) r\right]
 +{r}^{3}{\frac {df}{dr}} -f n{r}^{2} \right\}
\end{multline}
If we introduce a Regge-Wheeler ``tortoise'' coordinate $r^*$, such
that,
\begin{equation} \label{tortoi}
  \frac{d r^*}{dr} =\frac{1}{f(r)},
\end{equation}
an ``integrating'' factor $K(r(r^*))$,  and we also separate
variables
\begin{equation} \label{factorK1}
\Phi(t,r^*)= \frac{1}{K_V(r(r^*))} \phi(r(r^*)) e^{\w t},
\end{equation}
we find that the choice
\begin{equation} \label{factorK2}
K_V(r) =   \left[ r^n + \a (n-2) r^2 \frac{\p}{\p r}(r^{(n-3)}(\k-
f)) \right]^{-1/2}
\end{equation}
reduces (\ref{eqRW1}) to a (stationary) Schr\"odinger equation,
\begin{equation} \label{schro}
{\cal H} \phi \equiv  -\frac{\partial^2 \phi}{\partial {r^*}^2} +
V_V \phi  = -{\omega}^2 \phi \equiv E \phi
\end{equation}
with ``energy" eigenvalue $-\w^2$ and  ``potential'' $V(r(r^*))$
given by
\begin{equation} \label{poten1}
 V_V(r)= q_V(r) +\frac{1}{K_V(r)}\left[{{ \left( {\frac {d^{2}K_V}{d{r}^{2}}}
 \right)
  f^{2}}{}}+{
{ \left( {\frac {dK_V}{dr}}   \right) f {\frac {df}{dr}}
}{}}\right].
\end{equation}
The stability problem therefore reduces to analyzing the spectrum of
the Hamiltonian  operator (\ref{schro}), the presence of a negative
eigenvalue ($\w$ real) signaling an instability. ${\cal H}$  acts on
square integrable functions on the $f \geq 0$ region $I$, and, in
spite of the complexity of the potential $V_V$, information on its
spectrum can be obtained by
 using the $S-$deformation
approach, as done in \cite{koda,dge}. As in \cite{dge}, we find
that, due to the structure (\ref{poten1}) of  $V_V$, it  can be
$S-$deformed into $q_V$ given in (\ref{q1})-(\ref{q2}), thus, for
any normalized smooth test function of compact support in $I$
\cite{dge}
\begin{equation} \label{exH}
(\Phi,{\cal H} \Phi) = \int_{I}   \frac{|D \Phi|^2}{f} dr + \left(
k_V^2 - (n-1)\k \right)  \int_{I} \frac{|\Phi|^2  H}{r^2} \, dr,
\end{equation}
where $D \Phi = (\frac{d}{dr^*} + S) \Phi, S = - f d \ln(K_V)/dr$.
$E$ in (\ref{schro}) is greater than or equal to a lower bound  of
(\ref{exH}). Since neither $H$ nor $D$ depend on $k_V$, it is clear
that (\ref{exH}) can be made negative for sufficiently high $k_V$
unless $H$ is positive definite on $I$. In fact, if $H<0$ on an open
set $O \subset I$, the second integral in (\ref{exH}) will be
negative for a test function  with support contained in $O$, and
$(\Phi,{\cal H} \Phi)<0$ for high enough $k_V$. On the other hand,
if $H$ is nonnegative in $I$, then $E \geq 0$ since both integrals
in (\ref{exH}) will be positive and
\begin{equation}
k_V^2 - (n-1)\k > 0,
\end{equation}
as can be seen by integrating by parts
\begin{equation}
0 \leq \int_{\Sigma_{\k}^n} (\c^i V^j+\c^j V^i)(\c_i V_j+\c_j V_i)
\sqrt{\bar g} \; d^n x
\end{equation}
on the Riemannian compact manifold $\Sigma_{\k}^n$. We conclude that
a space-time is stable under vector perturbations if and only if $H$
in (\ref{q2}) is non negative on $I$. As a first application, note
from (\ref{q2}) that $H=1$ in Einstein theory ($\a=0$), an already
known \cite{koda} result on the stability of (\ref{b}) under vector
perturbations in General Relativity. Let us now consider  the
theories for which $P(\psi)$ in (\ref{p}) has two real roots
$\Lambda_1 < \Lambda_2$, so that $r$ extends to infinity (\ref{p}).
As in \cite{dge}, define
\begin{equation} \label{aux}
\psi_o = \frac{\Lambda_1+\Lambda_2}{2}, \hspace{1cm} \Delta =
\frac{\Lambda_1-\Lambda_2}{2}, \hspace{1cm} x =
\frac{\psi-\psi_o}{\Delta}
\end{equation}
in terms of which
\begin{equation} \label{h2}
H = {\frac {n-3}{2(n-1)}}+{\frac {n+1}{2 \left( n-1 \right)
{x}^{2}}}.
\end{equation}
Equation (\ref{h2}) shows that all solutions are stable for these
theories (note that that the Gauss-Bonnet term ${G_{(2)}}_b{}^a$ in
(\ref{lovelock}) is non trivial starting $n=3$, so that (\ref{h2})
is positive definite in the relevant cases). If $P(\psi)$ has
complex roots then $\Delta$ in (\ref{aux}) is purely imaginary and
$x^2 < 0$ in (\ref{h2}). For these theories the space-times
(\ref{b}) have a naked singularity \cite{dge}. The stability of
space-times with a naked singularity in EGB theories is currently
being studied \cite{dg3}.

\section{Scalar perturbations in EGB gravity}
\label{segb}

After interpolating the $n$ dependence in the perturbative equations
for $n \leq 9$, we obtained a set of equations equivalent to $\d
G_{ab}=0$, in terms of the functions introduced in (\ref{sca1}) and
(\ref{sgauge2}). The  $\delta G_{ij}=0$ equations imply the
condition
 \begin{eqnarray}
\label{deltaGij} 0 & = & \left( {\it F_{tt}}  -{\it F_{rr}}
 \right)  \left(  \left( n-2 \right)  \left( n-3 \right) \alpha\,
 \left( f  -\kappa \right) -{r}^{2}+ \left( n-2
 \right) \alpha\,r f'    \right)\nonumber \\
& & -2\,
 \left( n-2 \right)    \left( -{r}^{2}+
 \left( n-3 \right)  \left( n-4 \right) \alpha\, \left( f   -\kappa \right) +2\, \left( n-3 \right) \alpha\,r f'  +\alpha\,{r}^{2} f''
    \right) {\it H_L}.
\end{eqnarray}
\noindent  $\delta G_{rt}=0$ is equivalent to
 \begin{equation}
\label{deltaGrt} 0 =   n \,
  r \left( 2\,f -r f'  \right){\frac {\partial {\it H_L}}{\partial t}} -f    \left[n r
 {\frac {
\partial {\it F_{rr}}}{\partial t}}  +{k_S}^{2}
{\frac {
\partial {\it F_{rt}}}{\partial t}}   -2\,n{r}^{2}{\frac {\partial ^{2}{\it H_L}}{
\partial r\partial t}}     \right].
\end{equation}
\noindent  $\delta G_{tt}=0$ is equivalent to
 \begin{eqnarray}
\label{deltaGtt} 0 & = &
 \left[ {r}^{2}+ \left( n-1 \right)  \left( n-2 \right) \alpha\,
 \left( \kappa-f  \right)  \right]  \left[ n
f   r \left( -{\frac {\partial{\it F_{rr}} }{\partial r}}
  +2\,r{\frac {\partial ^{2}{\it H_L}}{\partial {r}^{2}}}
    \right)-2\, \left( n-1 \right) {\it H_L}    \left( {k_S}^{2}-n
\kappa \right) \right. \nonumber \\ & & \left. +nr \left( {\frac
{\partial {\it H_L}}{\partial r} }    \right)  \left( r f'
  +2\, \left( n-1 \right) f   \right)
-{\it F_{rr}}   \left( n \left( n-3 \right) f   +{k_S}^{2}+nr f'
\right)
   \right]
\nonumber \\ & &
 +nrf \left( r \right)  \left( -{\it F_{rr}}
   +2\,r{\frac {\partial {\it H_L}}{\partial r}}
    \right)  \left( 2\,r- \left( n-1 \right)  \left(
n-2 \right) \alpha\, f'    \right) \nonumber \\ & & +2\,
 \left( n-1 \right)  \left( n-2 \right) \alpha\,{\it H_L}
   \left( {k_S}^{2}-n\kappa \right)  \left( 2\,\kappa-2\,f
   +r  f'    \right).
\end{eqnarray}
\noindent $\delta G_{ir}=0,$ is equivalent to
\begin{eqnarray}
\label{deltaGir} 0 & = & 2 r f    \left[ 2\left( n-1 \right)  \left(
n-2 \right) \alpha
    \left( 2\,\kappa-2
f   +r f'    \right) {\frac {\partial {\it H_L}}{
\partial r}}
 +{\it
F_{rr}}    \left( 2\,r- \left( n-1 \right)  \left( n-2
 \right) \alpha f'    \right)  \right]
\nonumber \\ & & +\left[ {r}^{2}+ \left( n-1 \right)  \left( n-2
\right) \alpha\,
 \left( \kappa-f    \right)  \right]  \left[ 2\,r f
    \left( {\frac {\partial {\it F_{tt}}}{\partial r}}
   -2\, \left( n-1 \right) {\frac {\partial {\it H_L} }{
\partial r}}    \right) \right.
\nonumber \\ & & \left.
  +{\it F_{rr}}    \left( r f'   +2\, \left( n-3
 \right) f    \right) -2\, r {\frac {\partial^2 {\it F_{rt}}}{
\partial t^2}} +{\it F_{tt}}
  \left( -2\,f   +r f'   \right)  \right].
\end{eqnarray}
\noindent $\delta G_{it}=0,$ is equivalent to
\begin{eqnarray}
\label{deltaGit} 0 & = &
 \left(  \left( n-1 \right)  \left( n-2 \right) \alpha\, \left(
\kappa-f    \right) + {r}^{2} \right)  \left[ {\frac {\partial{\it
F_{rt}}}{
\partial t}}
   \left(  \left( n-4 \right) f   +r
 f'    \right) + {\frac {\partial^2{\it F_{rt}} }{
\partial r \partial t}}  f   r
-  {\frac {\partial {\it F_{rr}}}{\partial t}}
  r \right.
\nonumber \\ & & \left. -2\, \left( n-1 \right) r{\frac {\partial
{\it H_L}}{\partial t}}
     \right] +2\, \left( n-1 \right)  \left( n-
2 \right) r  {\frac {\partial {\it H_L}}{\partial t}}
    \alpha\, \left( r f'
-2\,f   +2\,\kappa \right) \nonumber \\ & &
 +   r f  \left( 2\,r- \left( n-1 \right)  \left(
n-2 \right) \alpha\, f'   \right){\frac {\partial{\it F_{rt}}}{
\partial t}}.
\end{eqnarray}
Finally, $\delta G_{rr}=0$ gives
\begin{eqnarray}
\label{deltaGrr} 0 & = &
 \left( {r}^{2}+ \left( n-1 \right)  \left( n-2 \right) \alpha\,
 \left( \kappa-f    \right)  \right)  \left\{ 2\,n{r}^{2}
{\frac {\partial ^{2}{\it H_L}}{\partial {t}^{2}}}  - f
  \left[ nr  {\frac {\partial{\it H_L} }{\partial r}}
    \left( r f'
  +2\, \left( n-3 \right) f    \right) \right. \right.
\nonumber \\ & & \left. \left. -nr
  {\frac {\partial {\it F_{tt}}}{\partial r}}
  f   -n{\it F_{rr}}    \left(
 \left( n-3 \right) f   +r f'
  \right) +2\,n  {\frac {\partial^2 {\it F_{rt}}}{\partial t^2}}  r-2\, \left( n-1 \right) {\it H_L}
    \left( {{k_S}}^{2}-n\kappa \right) +{{k_S}}^{2}{\it F_{tt}}
    \right] \right\}
\nonumber \\ & &
  - r f^2   \left( -{\it F_{rr}}   +2\,r{\frac {\partial {\it H_L}}{\partial r}}     \right)  \left( 2\,nr-n \left( n-1 \right)  \left( n-2
 \right) \alpha\, f'   \right)
\nonumber \\ & & -2\,
 \left( n-1 \right)  \left( n-2 \right) \alpha\, f {\it H_L}    \left( {{k_S}}^{2}-n\kappa \right)  \left( 2\,\kappa-2\,f
   +r f'   \right).
\end{eqnarray}

These equations are used to solve for the different variables, until
we get a single equivalent differential equation on a function of
$r$ and $t$, suitable for the stability analysis and other
applications discussed in the
 Conclusions. We first solve (\ref{deltaGij}) for $F_{tt}(t,r)$,

\begin{eqnarray}
\label{auxFtt}
  {\it F_{tt}}  & = & {\it F_{rr}}
\nonumber \\
& & + \,{\frac {2 \left( n-2 \right)      \left( -{r} ^{2}+\alpha\,
\left(  \left( n-3 \right)  \left( n-4 \right)  \left( f
   -\kappa \right) +2  f'  r \left( n-3 \right) +{r}^{2}
f''    \right)  \right) }{\alpha\, \left( n-2 \right)
 \left(  \left( n-3 \right)  \left( f  -\kappa
 \right) +r f'    \right) -{r}^{2}}}{\it H_L}
\end{eqnarray}
Integrating (\ref{deltaGrt}) with respect to $t$ gives
\begin{equation}
\label{auxFrr}
  {\it F_{rr}} ={\frac {{\it H_L}
   \left( -r f'   +2\,f
   \right) }{f   }}-{\frac {{{k_S}}^{2}{\it F_{rt}}
   }{nr}}+2\,r{\frac {\partial{\it H_L} }{\partial r}}
\end{equation}
plus an arbitrary function of $r$ that we absorb into $F_{rt}$ using
the freedom of adding to it an arbitrary function of $r$ (see
(\ref{sgauge2})). Inserting
\begin{equation}
\label{auxFFrt}
 {\it F_{rt}}  =: {\frac {r }{f  }}   \left( \Phi   +2\,{\it H_L}    \right)
\end{equation}
in (\ref{auxFrr}), and the resulting expression in (\ref{deltaGtt}),
we get an equation that can be solved  for $H_L$ in terms of $\Phi$,
and ${\partial\Phi }/{\partial r}$. The result is
\begin{eqnarray}
\label{auxHL}
  {\it H_L}  & = &
 \left[ \left( \alpha\, \left( n-1 \right)  \left( n-2 \right)  \left( f
   -\kappa \right) -{r}^{2} \right)  \left( n \left( -r
 f'   +2\,f    \right) -2\,
{{k_S}}^{2} \right) \right]^{-1} \nonumber \\
& & \times \left\{ n f r
 \left( \alpha\, \left( n-1 \right)  \left( n-2 \right)  \left( f
   -\kappa \right) -{r}^{2} \right) {\frac {\partial\Phi }{
\partial r}}   \right.
\nonumber \\  & & \left.
 +     \left[
 \left( n-1 \right)  \left( n-2 \right)  \left( f   n
 \left(  \left( n-3 \right)  \left( f  -\kappa
 \right) +r f'    \right) +{{k_S}}^{2} \left(
f   -\kappa \right)  \right) \alpha \right.\right. \nonumber \\  & &
\left. \left.
 -{r}^{2} \left( {{k_S}}^
{2}+n  f    \left( n-1 \right)  \right)  \right]\Phi \right\}
\end{eqnarray}
At this point
 we have explicit expressions for $F_{tt}$, $F_{rr}$, $F_{rt}$, and $H_L$,
all  in terms of $\Phi(t,r)$ and its first and second derivatives
with respect to
 $r$. Replacing these expressions in (\ref{deltaGir}) and (\ref{deltaGrr}) we end
 up with equations that contain ${\partial^3\Phi }/(\partial t^2\partial r)$, and
 ${\partial^3\Phi }/{\partial r^3}$. However, the linear combination
\begin{equation}
\frac{(-1/2)n f {\cal A} + {\cal B}}{n{r}^{2} \left( n-1 \right)
\left( n-2 \right)  \left( f \left( r
 \right) -\kappa \right) \alpha-n{r}^{4}},
\end{equation}
where ${\cal A}$ and ${\cal B}$ are the RHSs of (\ref{deltaGir}) and
 (\ref{deltaGrr}) respectively, eliminates both
 terms, giving an equation of the form,
\begin{equation}
\label{equPhi} {\frac{\partial^2\Phi }{\partial t^2}}-f^2
{\frac{\partial^2\Phi }{\partial r^2}} +
 Q_S {\frac{\partial\Phi }{\partial r}} + q_S \Phi = 0.
\end{equation}
where $Q_S$ and $q_S$ are functions of $r$. One can show that any
solution of (\ref{equPhi}), when inserted back
into the formulas for $F_{tt},F_{rr},F_{rt}$ and $H_L$ in terms of
$\Phi$, solves all  the $\d G_{ab}=0$ equations. Note that
(\ref{equPhi})  is of the same form as (\ref{eqRW1})
 for the vector perturbations. However  $Q_S$ and $q_S$ in (\ref{equPhi})
depend not only on $r$, but also on {\em all} the parameters of the theory in a rather
complicated
 way. In particular, we do not find the ``factorization'' property for the
 dependence on ${k_S}^2$, neither we find that $q_S(r)$ has a definite sign.
 Therefore, the S-deformation method does not appear to be readily applicable for
 the stability analysis in this case. On the other hand, and perhaps a little
 surprisingly, we have found an explicit form for the transformation that puts
(\ref{equPhi})
 in a standard Regge-Wheeler-Zerilli form.
Namely, we  introduce a function $\hat \phi(t,r)$ such that
\begin{equation}
\label{auxchi} \Phi(t,r)=K_S(r)\hat \phi(t,r).
\end{equation}
Then, choosing the integration factor
\begin{equation}
\label{auxKsca}
  {\it K_S} \left( r \right) ={\frac {{r}^{-1/2\,n+1} \left( nr
 f'   -2\,f   n+2\,{{k_S}}^{2}
 \right) }{\sqrt {{r}^{2}- \left( n-2 \right) \alpha\,\left[r
 f'  - \left( n-3 \right)
 \left( \kappa-f    \right)\right] }}}
\end{equation}
and switching to  tortoise coordinate $r^*$  (\ref{tortoi}) cancels
the terms in $\partial \hat \phi(t,r) /\partial r^*$ and yields
 an equation of the Regge-Wheeler-Zerilli form, which, after
 separating variables $\hat \phi(t,r) = \phi(r) e^{\w t}$
gives (as in (\ref{schro})) a stationary Schr\"odinger equation
\begin{equation}\label{ss}
{\cal H} \phi \equiv  -\frac{\partial^2 \phi}{\partial {r^*}^2} +
V_S \phi  = -{\omega}^2 \phi \equiv E \phi
\end{equation}
Unfortunately, the explicit expression for the scalar ``potential''
$V_S$
 in terms of
 $r^*$ and the parameters of the static solution is extremely long and complicated,
 and we have not been able to put it in a form  that would be useful for a general
 analysis of the stability problem. It should be clear, nevertheless, that,
 for any choice of parameters, including $n$, $V_S$ can be straightforwardly
 recovered, e.g., by means of a symbolic manipulation program, following the
 procedure outlined above. In the next subsection, some
 examples are analyzed by assigning particular values to the parameters.

\subsection{Application: a scalar instability in small mass, $5D$
spherical EGB black holes}

For $\Lambda=0$ and $\k=1$, the $\epsilon=-1$ branch of (\ref{f1}),
\begin{equation} \label{fs}
 f(r) = 1 +\frac{r^2}{\a(n-1)(n-2)} \left( 1 -
 \sqrt{1+\frac { 4 \a \m \left( n-1 \right)  \left( n-2 \right)}{n r^{n+1}}
  } \; \right),
\end{equation}
reduces to the $n+2$ dimensional Schwarzschild-Tangherlini \cite{st}
(Einstein) black hole in the $\a \to 0$ limit:
\begin{equation} \label{fst}
f(r) = 1 -\frac{2 \m}{n r^{n-1}} + {\cal O} (\a)
\end{equation}
In this section  we will apply our  results to exhibit a low mass
scalar instability of 5D Schwarzschild-Tangherlini-EGB black holes.
\\

Before that, we note that, in general,  $\a$ (assumed strictly
positive from now on) can be used to introduce dimensionless
quantities
\begin{equation}
\mm := \frac{\m}{\a^{(n-1)/2}} \hspace{1cm} \ll := \L \a
\hspace{1cm} x := r \a^{-1/2}
\end{equation}
and that $f$ in (\ref{f1}) depends on $\a$ only through   $x,\mm$
and $\ll$. If we  define
 \begin{equation} \label{xs}
 x^*:= \frac{r^*}{\sqrt{\a}} = \int_{x_o}^x \frac{dx'}{f(x',\mm,\ll)}
 \end{equation}
 we find that (\ref{ss}) is equivalent to
\begin{equation}
-\frac{\partial^2 \phi}{\partial {x^*}^2} + \a V_S \phi  = \a E
\phi.
\end{equation}
Furthermore, it can be shown that $\a V_S = \tilde
V(x,\mm,\ll,k_S{}^2)$, so that the stability problem reduces to
determining if the  $\a-$independent potential $\tilde
V(x,\mm,\ll,k_S{}^2)$ admits negative energy eigenvalues. For $n=3$
ST-EGB black holes
\begin{equation} \label{f3}
f(x,\mm) = 1 + \frac{x^2}{2} \left[ 1 - \sqrt{1 + \frac{8 \mm}{3
x^4}} \; \right]
\end{equation}
 and we must assume $\mm > 3/2$, since  there is
no horizon below this mass value. This is a special feature of $n=3$
EGB, for higher dimensions there is always a horizon \cite{dge}. The
integral (\ref{xs}) for $f$ given in (\ref{f3}) can be solved in
closed form in terms of hypergeometric functions, and values of
$x^*$ and $\tilde V$ can be obtained for different $x's$ and used to
generate parametric plots of $\tilde V$ $vs.$ $x^*$.
 All graphs in figures 1-6
were generated setting $x_o=2x_H$ in (\ref{xs}), so that $x^*=0$
when $x=2x_H$, and $x^* \to -\infty$ as $x \to x_H{}^+$.  Since
$\tilde V (x^*)$ is bounded and $ \lim _{x^* \to \pm \infty} V
(x^*)=0$, a sufficient condition for the existence of a bound state
of negative energy is \cite{ajp}
\begin{equation} \label{cond}
\int_{-\infty}^{\infty}  V \; dx^* < 0.
\end{equation}
The above condition is certainly met  in the graphs shown of figures
1 and 2, which exhibit the $\ell=2$ and $\ell=10$ potentials for
$\mm=1.7$ \footnote{A perturbation obtained from the $\ell=0$ scalar
harmonic correspond to a variation of the mass parameter, whereas an
$\ell=1$ perturbation  is shown to be pure gauge \cite{koda}. The
first relevant case is therefore $\ell=2$.}. For higher mass values
(\ref{cond}) may not be satisfied, and still the potentials may
allow  negative energy bound states. This is illustrated in figures
3-5. For these examples, we have fit a quadratic curve around the
potential minimum, and checked that the ground state gaussian wave
function $\phi$  of the associated harmonic oscillator (plotted
together with $\a V_S$) has a negative energy expectation value
\begin{equation}\label{nev}
\int_{-\infty}^{\infty} \phi^* \left( -\frac{\partial^2
\phi}{\partial {x^*}^2}  + \a V_S \right) \phi \; dx^* < 0.
\end{equation}
This shows that the spectrum contains negative energy eigenfunctions
and that the black hole is unstable.  For higher mass values in
$n=3$ (i.e, spacetime dimension five), as well as for higher
spacetime dimensions, we were not able to find test functions with
negative ``energy" expectation values, there seems to be no scalar
instability in these cases (see, e.g., Figure 6), although a more
systematic study has not been completed yet. Note however that  a
low mass instability under {\em tensor} mode perturbations was
previously found in six dimensional Schwarzschild-Tangherlini-EGB
black
holes \cite{dgl,dge}. \\

\section{Conclusions}\label{conc}

The study of the  linear stability of static solutions to the
Einstein-Gauss-Bonnet gravity with spatial slices of the form
$\Sigma_{\k}^n \, \times \, {\mathbb R}^+$, $\Sigma_{\k}^n$ an
$n-$manifold of constant curvature $\k$ can be carried out using
the techniques introduced by Kodama and collaborators. In the
classification of Kodama, et. al. \cite{koda}, general linear
perturbations can be constructed from appropriate harmonic tensors
on $\Sigma_{\k}^n$, and classified accordingly into tensor, vector
and scalar modes. Tensor perturbations were analyzed in
\cite{dgl,dge}, the other modes being the subject of this paper. We
proved that, as happens  in higher dimensional GR stability problems
\cite{koda,gh}, the perturbation equations can be  reduced to a
single stationary Schr\"odinger-like equation on a function of the
radial coordinate, the potential being different for the scalar and
vector cases. Finding the linearized EGB equation and the
appropriate integrating factors leading to the equivalent
Schr\"odinger problem is far more difficult in the vector and scalar
modes than in the tensor modes. However, after performing the
calculations in spacetime dimensions five to eleven, the dimension
dependence of the equations was interpolated by formulas that we
believe are valid beyond this range and that  reproduce the expected
$\a \to 0$ (GR) results. \\
In the vector case, we were able to prove stability using an
S-deformation argument. The same result is found in higher
dimensional GR. In the scalar case, however, we found an instability
that is absent in GR \cite{koda,gh}. Although the complexity of the
potential prevents us from drawing general conclusions, the EGB
analogous to 5D Schwarzschild black holes are shown to be unstable
below a critical mass. A similar result was found in \cite{dgl},
where it was proved that low mass 6D spherical, asymptotically
Euclidian black holes are unstable.  These results are relevant to
TeV scale quantum gravity scenarios,  where those black holes are
predicted to be produced in high energy collisions
\cite{lhc}.\\
 The
potentials that we have obtained have a number of applications
beyond the study of the stability of black holes and cosmological
solutions. Among the most immediate ones are the analysis of black
hole uniqueness, quasi normal modes \cite{gm},  and  the analysis of
stability of naked singularities in EGB gravity. These topics are
currently under study \cite{dg3}.

\section*{Acknowledgments}

This work was supported in part by grants of the Universidad
Nacional de   C\'ordoba and Agencia C\'ordoba Ciencia (Argentina).
It was also supported in part by grant NSF-INT-0204937 of the
National Science Foundation of the US. The authors are supported by
CONICET (Argentina).

\pagebreak

\begin{figure}[h]
\includegraphics[width=4in]{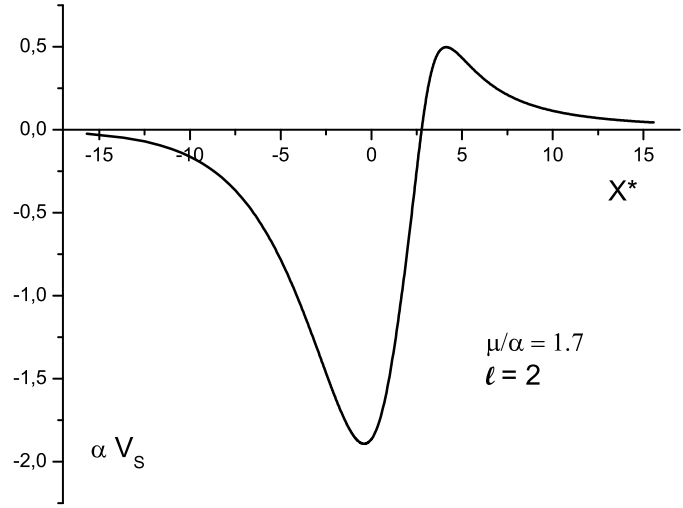}
\caption{\label{fig4} The potential $\tilde V = \a V_S$ as a
function of $x^* = r^*/\sqrt{\a}$, for $n=3$, $\Lambda=0$, $\tilde
\mu= \m / \a = 1.7$. The scalar perturbation corresponds to the
$\ell=2$  harmonic.}
\end{figure}

\begin{figure}[h]
\includegraphics[width=4in]{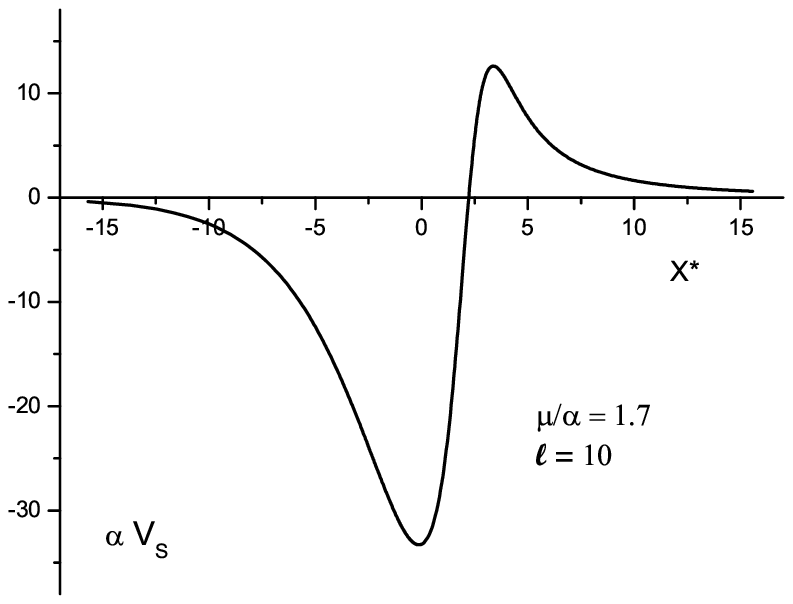}
\caption{\label{fig4} The potential $\tilde V = \a V_S$ as a
function of $x^*= r^*/\sqrt{\a}$, for $n=3$, $\Lambda=0$, $\tilde
\mu= \m / \a = 1.7$. The scalar perturbation corresponds to the
$\ell = 10$ harmonic.}
\end{figure}

\begin{figure}[h]
\includegraphics[width=4in]{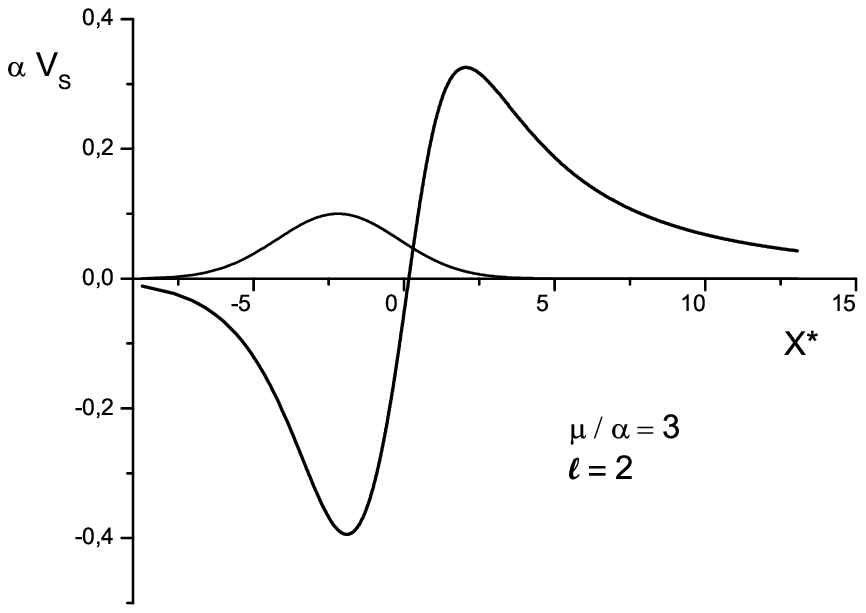}
\caption{\label{fig4}The potential $\tilde V = \a V_S$ and a (non
normalized) gaussian test wave function as a function of $x^*=
r^*/\sqrt{\a}$, for $n=3$, $\Lambda=0$, $\tilde \mu= \m / \a = 3$.
The scalar perturbation corresponds to the $\ell=2$  harmonic. The
normalized  test function gives $< -d^2/d{x^*}{}^2 > = 0.12$ and $<
\a V_S > \simeq -0.28$. The expectation value of the ``Hamiltonian"
is negative for this test function, implying the existence of
negative energy eigenvalues.}
\end{figure}

\begin{figure}[h]
\includegraphics[width=4in]{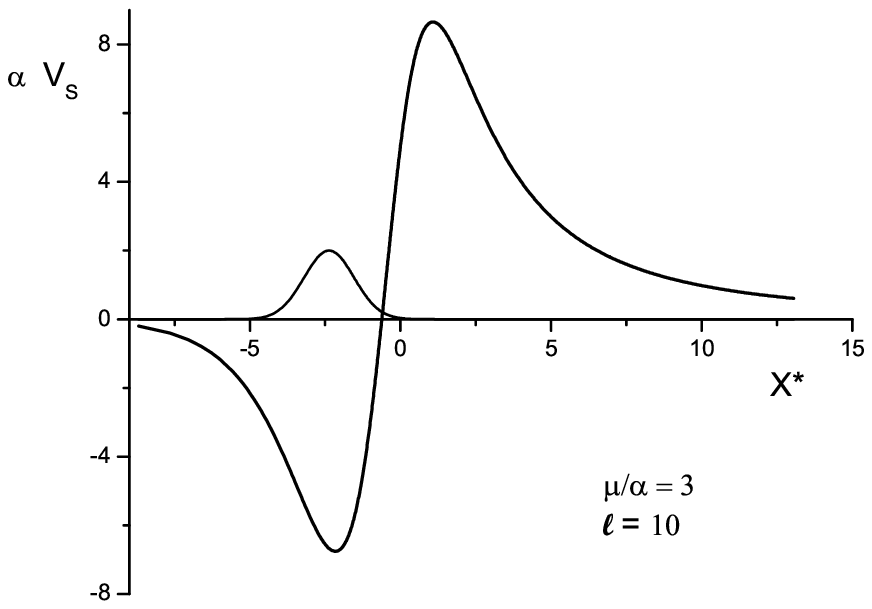}
\caption{\label{fig4} The potential $\tilde V = \a V_S$ and a (non
normalized) gaussian test wave function as a function of $x^*=
r^*/\sqrt{\a}$, for $n=3$, $\Lambda=0$, $\tilde \mu= \m / \a = 3$.
The scalar perturbation corresponds to the $\ell=10$  harmonic. The
normalized  test function gives $< -d^2/d{x^*}{}^2 > = 0.73$ and $<
\a V_S > \simeq -6.11$. The expectation value of the ``Hamiltonian"
is negative for this test function, implying the existence of
negative energy eigenvalues.}
\end{figure}

\begin{figure}[h]
\includegraphics[width=4in]{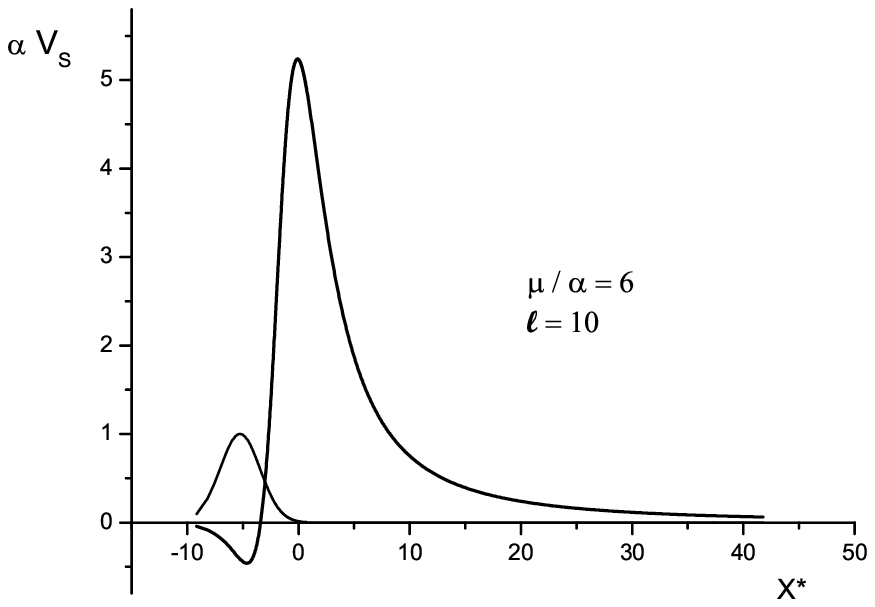}
\caption{\label{fig5} The potential $\tilde V = \a V_S$ and a (non
normalized) gaussian test wave function as a function of $x^*=
r^*/\sqrt{\a}$, for $n=3$, $\Lambda=0$, $\tilde \mu= \m / \a = 6$.
The scalar perturbation corresponds to the $\ell=10$  harmonic. The
normalized  test function gives $< -d^2/d{x^*}{}^2 > = 0.15$ and $<
\a V_S > \simeq -0.24$. The expectation value of the ``Hamiltonian"
is negative for this test function, implying the existence of
negative energy eigenvalues.}
\end{figure}

\begin{figure}[h]
\includegraphics[width=6in]{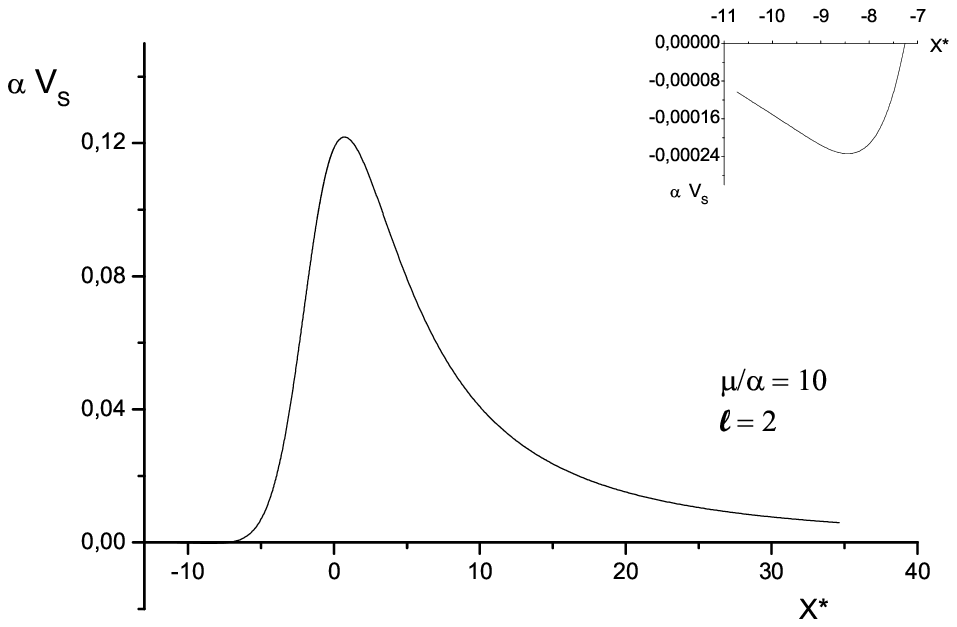}
\caption{\label{fig6} The potential $\tilde V = \a V_S$ as a
function of $x^*= r^*/\sqrt{\a}$, for $n=3$, $\Lambda=0$, $\tilde
\mu= \m / \a = 10$. The scalar perturbation corresponds to the $\ell
= 2$  harmonic. $\a V_S$ has a small negative tail -shown in the
inset- which cannot accommodate a negative energy state.}
\end{figure}


\begin{thebibliography}{99}


\bibitem{str1} D. J. Gross and E. Witten, Nucl. Phys. {\bf B277}, 1 (1986);
B. Zumino, Phys. Rep. {\bf 137}, 109 (1985); B. Zwiebach, Phys.
Lett. {\bf 156B}, 315 (1985); D. Friedan, {\it Phys. Rev. Lett.}
{\bf 45} (1980) 1057; I. Jack, D.Jones and N. Mohammedi, Nucl. Phys.
{\bf B322}, 431 (1989); C. Callan, D. Friedan, E. Martinec and M.
Perry, Nucl. Phys. {\bf B262}, 593 (1985).
\bibitem{Lovelock} D. Lovelock, {\it Jour. Math. Phys.} {\bf 12} (1971) 498.
\bibitem{jm} J.  Maldacena,  {\it Adv. Theor. Math. Phys.} {\bf 2} (1998) 231.
\bibitem{w1} J. T. Wheeler,  {\it Nuc. Phys.} {\bf B268} (1986) 737
\bibitem{w2} J. T. Wheeler, {\it Nuc. Phys.} {\bf B273} (1986) 732.
\bibitem{bd} D. G. Boulware
and S. Deser, {\it Phys. Rev. Lett.} {\bf 55}(1985) 2656.
\bibitem{w3} B. Whitt, {\it Phys. Rev.} {\bf D38} (1998) 3000.
\bibitem{atz} Aros R, Troncoso R and  Zanelli J 2001 {\it Phys. Rev.}
 {\bf D63}084015
 \bibitem{chm} Rong-Gen Cai,
Phys. Rev. D 65, 084014 (2002);  {\it Phys. Lett.} {\bf B 582} 237
(2004).
\bibitem{koda} H. Kodama, A. Ishibashi and O. Seto, Phys. Rev. {\bf D62} (2000)
064022.\\H. Kodama and A. Ishibashi {\it Prog.Theor.Phys.} {\bf 111}
(2004) 29, hep-th/0308128;
 H. Kodama and A. Ishibashi,  gr-qc/0312012.
\bibitem{dgl} G. Dotti and R. J. Gleiser, {\it Class. Quantum Grav.} {\bf 22}
(2005), L1.
\bibitem{dge} G. Dotti and R. J. Gleiser, {\it Phys. Rev} {\bf D 72},
044018 (2005), gr-qc/0503117.
\bibitem{higu} A. Higuchi,{\it Jour. Math. Phys.} {\bf 28}
(1987) 1553 (Erratum ibid.43:6385,2002).
\bibitem{st}  F.R. Tangherlini, {\it Nuovo Cimento} {\bf 27}, 365
(1963).
\bibitem{ajp} W. F. Buell and B. A. Shadwick {\it Am. J. Phys.} {\bf 63} (1995),
256.
\bibitem{gh} G. Gibbons and S. Hartnoll, {\it Phys. Rev.} {\bf D66} (2002)
064024.
\bibitem{lhc} S. Dimopoulos, G. Landsberg,  {\it Phys.Rev.Lett.} {\bf 87}
161602 (2001),  hep-ph/0106295.
\bibitem{gm} R. A. Konoplya, Phys. Rev. D 71, 024038 (2005);
Phys. Rev. D 68, 124017 (2003); Phys. Rev. D 68, 024018 (2003), E.
Abdalla, R.A. Konoplya, C. Molina, {\it
    Phys.Rev.} {\bf D72 }084006, (2005), hep-th/0507100.
\bibitem{dg3} G.Dotti and R.J.Gleiser, work in progress.

\end{thebibliography}
\end{document}